\documentclass[aps, superscriptaddress,  preprint,amsfonts, amsmath, amssymb, floatfix]{revtex4-2}

\usepackage{graphicx}
\usepackage{float}
\usepackage{sidecap}
\usepackage{dcolumn}
\usepackage{bm}
\usepackage{epsfig}
\usepackage{braket}
\usepackage{color}
\usepackage[toc,page]{appendix}
\usepackage[colorlinks=true, pdfstartview=FitV, linkcolor=blue, citecolor=blue, urlcolor=blue]{hyperref}

\begin{document}
\title{A Programmable Spatiotemporal Quantum Parametric Mode Sorter}

\author{Malvika Garikapati}
\thanks{These two authors contributed equally}
\author{Santosh Kumar}
\thanks{These two authors contributed equally}
\email{skumar5@stevens.edu}
\author{He Zhang}
\author{Yong Meng Sua}
\author{Yu-Ping Huang}
\email{yhuang5@stevens.edu}

\affiliation{Department of Physics, Stevens Institute of Technology, Hoboken, NJ, 07030, USA}
\affiliation{Center for Quantum Science and Engineering, Stevens Institute of Technology, Hoboken, NJ, 07030, USA}

\date{\today}



\begin{abstract}
We experimentally demonstrate a programmable parametric mode sorter of high-dimensional signals in a composite spatiotemporal Hilbert space through mode-selective quantum frequency up-conversion. As a concrete example and with quantum communication applications in mind, we consider the Laguerre-Gaussian and Hermite-Gaussian modes as the spatial and temporal state basis for the signals, respectively. By modulating the spatiotemporal profiles of the up-conversion pump, we demonstrate the faithful selection of single photons in those modes and their superposition modes. Our results show an improvement in the quantum mode-sorting performance by coupling the up-converted light into a single-mode fiber and/or operating the upconversion at the edge of phase matching. By optimizing pump temporal profiles only, we achieve more than 12 dB extinction for mutually unbiased basis (MUB) sets of the spatiotemporal modes. This fully programmable and efficient system could serve as a viable resource for quantum communications, quantum computation, and quantum metrology.

\end{abstract}

\maketitle

\section*{INTRODUCTION}
Nowadays, quantum parametric frequency conversion processes such as spontaneous parametric down-conversion \cite{scully_zubairy_1997, SPDC,Rozenberg:22}, second harmonic generation \cite{Chen:19}, sum/difference-frequency generation, and four-wave mixing \cite{Lu2019} are extensively utilized in quantum optics \cite{Ghosh87, Kumar:90, Hochrainer2022}, with applications in hybrid quantum networks \cite{PRApp18, Niizeki2020}, creating entanglement between disparate quantum memories \cite{Albrecht2014, Bock2018}, realizing low-noise infrared upconversion photon counters \cite{Dam2012,Israelsen:21,Huang2022}, and so on. 
Meanwhile, quantum systems subtending high-dimensional (HD) Hilbert spaces have been studied due to their advantages in a higher information capacity, enhanced security, or increased resistance to noise. Thus far, higher-dimensional information coding has been demonstrated on various platforms such as those of trapped ions \cite{PhysRevX.5.021026}, polar molecules \cite{Yan2013}, Rydberg atoms, cold atomic ensembles \cite{Parigi2015}, photonic systems, or superconducting phase qudit \cite{doi:10.1126/science.1173440}. In photonic systems, the information can be encoded in an HD Hilbert space with multiple degrees of freedom (DoF) over polarization, spatial, time, energy, and optical path \cite{Wang2015,Brecht15,Boucher:20,Ploschner2022,Wright:22,He2022}. This multiple DoF of the photons can be coherently coupled with each other to boost quantum communication capabilities \cite{Erhard2018} with broad applications in quantum key distribution \cite{Raymer:20, Otte:20}, quantum logic operations, quantum teleportation, quantum metrology \cite{Habif:21,PRXQuantum_21,PRL_22} and so on. 

A challenge to HD quantum information processing is with efficient sorting and processing of the highly multiplexed modes. Indeed, several techniques have been demonstrated to sort the spectral modes by using linear-optical resonators \cite{PRA_Shuang18,Wei2020}, and sort the spatial \cite{Mair2001,PRL_Berkhout,Fontaine2019,zhang_mode_2019,Defienne_2020,Pinnell:20} and temporal \cite{Eckstein:11,shahverdi_quantum_2017,Raymer:20, Reddy:13} modes, respectively, by using nonlinear optics. Among them, those nonlinear-optical techniques have an upper hand in mode sorting as compared to their linear counterparts, as they can manipulate single photons without any loss.

While most of the above work has focused on sorting individual modes, the ability to sort combined spatial-temporal (ST) modes could find vast applications in developing practical quantum technologies \cite{Yorulmaz:14,Manurkar:16,PhysRevX.10.031031,Raymer_2020,Ploschner2022}. Recently, we demonstrated an HD quantum system by combining spatial and temporal DoF by using a spatial light modulator (SLM) and optical delay line (ODL), respectively \cite{Kumar_2021}. There, we demonstrated efficient mode sorting by selectively upconverting signal modes with spatially modulated and temporally delayed pumps. It has a limited capacity due to the delay-only temporal control of the Gaussian pulses, which restricts its advantage for HD Hilbert space quantum communication and computation. 

In this article, we overcome the aforementioned restriction of our previous study by adding arbitrary temporal modulation and demonstrating HD quantum parametric mode sorting (QPMS) over the compound ST Hilbert space with arbitrarily re-configurable modes. Specifically, we utilize spatial light modulators for controlling spatial DoF \cite{Santosh19, Kumar_2021} and an optical arbitrary waveform generator (OAWG) based on spectral line-by-line pulse shaping to manipulate the amplitude and phase profiles of the time-frequency DoF \cite{PhysRevLett.94.073601,shahverdi_quantum_2017, Jiang:07}. We experimentally demonstrate that the mutually unbiased basis (MUB) in an HD Hilbert space can be selectively up-converted according to their spatial and/or time-frequency modes. This allows photon-efficient applications over overlapping yet orthogonal spatial and temporal modes, thus accessing a much larger Hilbert space than allowed by each individual DoF \cite{Brecht15,PhysRevX.10.031031,Cruz-Delgado2022}. This also helps to achieve high selectivity by merging the spatial and time-frequency modes, which will reduce error probabilities in HD quantum applications. With this efficient and programmable manipulation of HD ST modes, it could find practical applications in emerging quantum technologies, including robust HD quantum key distribution.

\section*{MODEL}
\begin{figure*}[!ht]
   \includegraphics[width=1\textwidth]{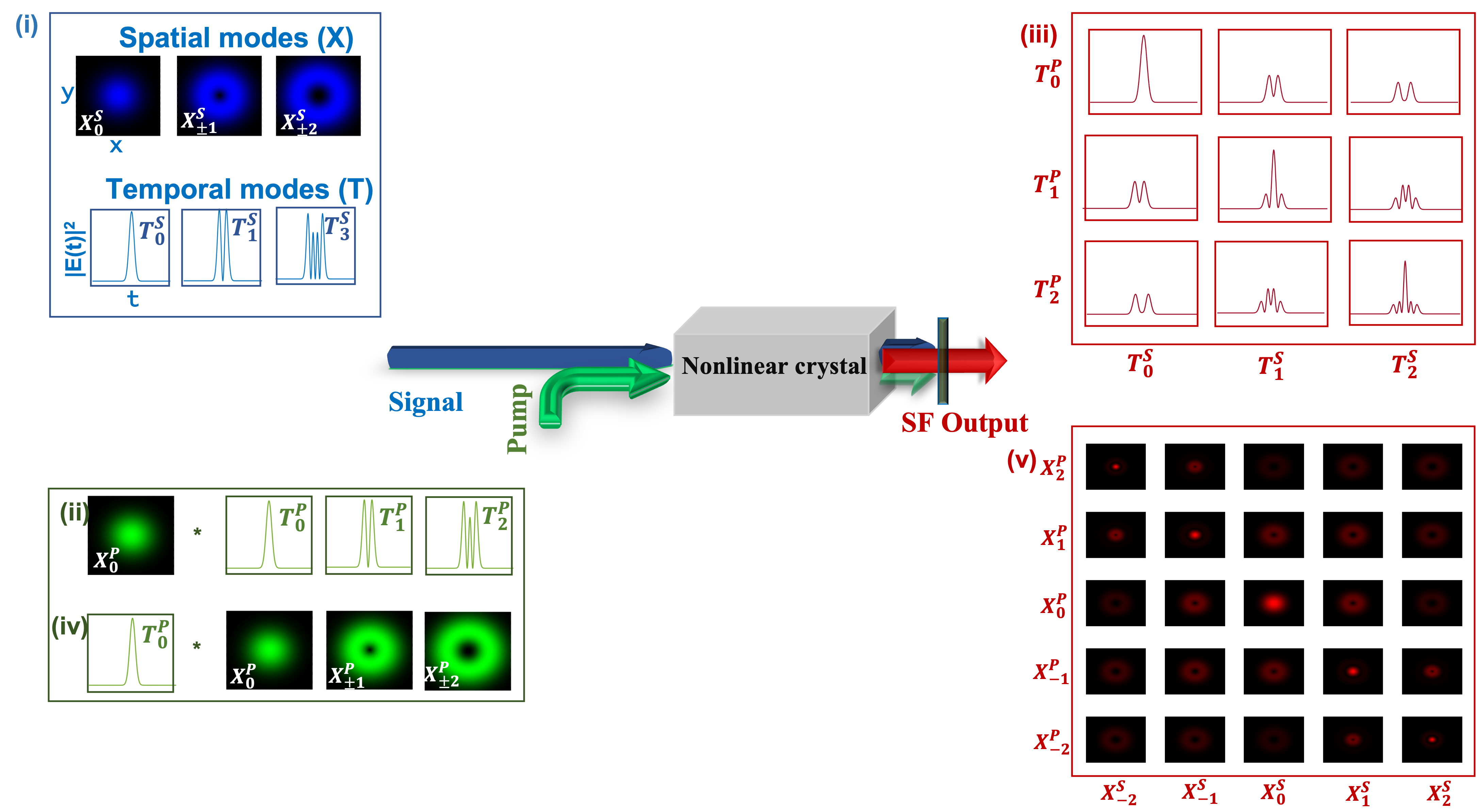}
    \caption{Schematic of ST quantum parametric mode sorter using a $\chi^2$ nonlinear crystal. (i) On the top left represents the signal states in blue, whereas, the bottom left states in green represents the pump states. (ii) represents a set of pump modes with a Gaussian spatial profile $|X^p_0\rangle$ and different temporal profiles $|T^p_0\rangle$, $|T^p_1\rangle$, and $|T^p_2\rangle$. Whereas (iv) represents pump states with a Gaussian temporal profile $|T^p_0\rangle$ and different spatial profile $|X^p_{-2}\rangle$, $|X^p_{-1}\rangle$, $|X^p_{0}\rangle$, $|X^p_{1}\rangle$ and $|X^p_{2}\rangle$. The states in red on the right represent the SF output profiles. (iii) represents SF temporal profiles when the signal is combined with the pumps of (ii). Similarly, (v) represents the output SF spatial profiles for pumps of (iv).}
    \label{fig:Model}
\end{figure*}

We consider photons in ST modes $\Psi = E_r(x,y)E_t(t)$, where $E_r(x,y)$ and $E_t(t)$ are the electric fields in the spatial and temporal domains, respectively. We exploit an HD Hilbert space by preparing quantum states in Laguerre-Gaussian (LG) spatial mode basis and Hermite-Gaussian (HG) temporal basis. In order to retrieve relevant quantum information, we need to extract the desired mode among a plethora of modes generated by processes like spontaneous four-wave mixing \cite{Lu2019, Clemmen:09, Sharping:06}, multimode spontaneous parametric down-conversion \cite{2012EPJD...66..263M, huang_mode-resolved_2013,PhysRevX.10.031031}, and antibunching photon emission in quantum dots \cite{Santori:03, Cui:05}. We present a QPMS scheme that is realized through sum frequency (SF) generation in a second-order $\chi^2$ nonlinear crystal described by a set of coupled differential equations (see \cite{Kumar_2021} for details). In our current setup, we use a pump and signal input pulses having similar wavelengths under the undepleted pump approximation where the inverse group velocity mismatch between them can be neglected, so that our system operates in the single side-band velocity mismatch (SSVM) regime \cite{Reddy:13, huang_mode-resolved_2013}. When the temporal pulse width of the pump is smaller than the temporal walk-off between the pump and SF pulses in the crystal, high mode selectivity can be achieved. We numerically simulate the results using an adaptive split-step Fourier method under the slowly varying envelope approximation. For both simulation and experiment discussions, we define the selectivity ($S$) for each quantum state of the pump as: 
\begin{equation} \label{eq:ER}
    S = 10log_{10}\left(\frac{N_D}{\sum_{i\neq D} N_i}\right),
\end{equation}
where $N_D$ is the number of SF photons in the `Desired' mode among the other considered modes, as measured using the Si avalanche photo-diode (Si-APD). Figure \ref{fig:Model} shows a schematic of our QPMS system that can selectively up-convert the desired signal mode among overlapping LG spatial modes and HG temporal modes by passing it through a nonlinear crystal with a specifically designed ST pump. 

In this paper, we exclusively consider spatial LG modes (azimuthal index mode $l$ and zero radial indexes), hence simplifying the notation to $|X_{l}^{p,s}\rangle$ where `$l$' represents the orbital angular momentum mode number (0, $\pm$1, $\pm$2,...), `$p$' stands for pump and `$s$' stands for signal pulses. Similarly, we consider HG temporal modes that are represented by $|T_{m}^{p,s}\rangle$ where `$m$' is the order of HG mode (0,1,2,...). We consider that the information is stored in a unique quantum state among 3 HG temporal modes $|T_0\rangle$, $|T_1\rangle$, $|T_2\rangle$ and 5 LG spatial modes $|X_{-2}\rangle$, $|X_{-1}\rangle$, $|X_{0}\rangle$, $|X_{1}\rangle$, $|X_{2}\rangle$. To selectively up-convert the signal photons in each of these 15 ST modes, we use a pump with a unique ST mode profile. We monitor and analyze the SF output for each of the 225 pairs of the ST pump and signal modes.

To thoroughly understand our hybrid system, we first illustrate the outcome of spatial and temporal processes independently. In Fig. \ref{fig:Model} (i), we prepare spatial and temporal quantum states for the signal. In Fig. \ref{fig:Model} (ii) we prepare pump states with different temporal modes and a constant spatial mode. We employ a Gaussian configuration for the spatial DoF to achieve maximum conversion efficiency and clearly recognize the features of the temporal DoF. Fig. \ref{fig:Model} (iii) represents the simulated SF outcome for a combination of these states. It is constructed by combining specific $|T_{1, 2, 3}\rangle$ modes for both pump and signal, and calculating the SF output at the different delays. We consider the SF counts at zero delays to accurately estimate the SF counts at the central overlap. We observe high SF counts along the diagonal elements which correspond to identical pump and signal temporal modes as $|T_{0}^{p}, T_{0}^{s}\rangle$, $|T_{1}^{p}, T_{1}^{s}\rangle$ and $|T_{2}^{p}, T_{2}^{s}\rangle$. The off-diagonal elements correspond to SF generation for pump and signal with orthogonal temporal modes like $|T_{0}^{p}, T_{1}^{s}\rangle$, $|T_{0}^{p}, T_{2}^{s}\rangle$, $|T_{1}^{p}, T_{0}^{s}\rangle$, $|T_{1}^{p}, T_{2}^{s}\rangle$, $|T_{2}^{p}, T_{0}^{s}\rangle$ and $|T_{2}^{p}, T_{1}^{s}\rangle$. These modes show significantly lower SF output at zero delays. The spectral profiles of 2ps signal and pump temporal pulses are shown in Appendix A. Similarly, in Fig. \ref{fig:Model} (iv) we prepare different spatial modes by maintaining a constant temporal mode with a Gaussian profile. We plot the intensity and shape of the output SF pulses in Fig. \ref{fig:Model}. (v) by simulating spatial $|X_{i}\rangle$ modes for both input pulses and calculating the output at each spatial location after passing it through the same crystal. We observe higher intensity for the diagonal elements as compared to off-diagonal elements. Since the diagonal terms correspond to pump mode $|X_{i}^{p}\rangle$ which efficiently upconverts signal modes $|X_{-i}^{s}\rangle$ \cite{zhang_mode_2019}. We also experimentally capture these spatial SF output modes on a CCD camera, as shown in Appendix B.

We further optimize the pump temporally using particle swarm optimization (PSO) to achieve better fidelity in selecting the desired signal. Likewise, we use this technique to simulate the arbitrary shape of the pump according to the experimental condition (as discussed in the next section). Here, we iteratively optimize the selectivity by individually manipulating the relative phase of 37 evenly spaced frequency comb lines simultaneously on an ensemble of 16 potential candidates (C). The phase change of each comb line in a particular candidate is influenced by its last known phase, its best-known phase (Personal Best), and the best-known phase of the entire ensemble (Global Best) with an adaptively changing weight component as shown below:
\begin{equation} \label{eq:PSO}
    C_n^{(i+1)} = w * C_n^{(i)} + w_p * R_1 * C_{PB}^{(i)} + w_g * R_2 * C_{GB}^{(i)},
\end{equation}
where $w$, $w_p$ and $w_g$ are the adaptively selected weight parameters, $C_{PB}^{(i)}$ and $C_{GB}^{(i)}$ are the personal and global best candidates achieved up to the $i^{th}$ iteration. $R_1$ and $R_2$ are randomly generated weights parameters added to utilize the search space intelligently. To further improve the selectivity, we can similarly optimize the pump in the spatial domain, as discussed in our previous work \cite{Santosh19}.

\section*{EXPERIMENTAL SETUP}
A continuous wave light from a tunable laser source (TLS) at the central wavelength, 1546.6 nm, is sent to the optical frequency comb generator (OFCG). We used a commercially available optical frequency comb generator (Optocomb, WTEC-01-25). It consists of a phase modulator in a Fabry-Pivot cavity driven by a 25 GHz radio frequency signal and generates a phase-coherent broadband frequency comb with a line spacing 25GHz. The output of the OFCG is low ($\sim$ -20dBm) which is then amplified by erbium-doped fiber amplifiers (EDFAs). The amplified comb lines are manipulated individually (line-by-line) in amplitude and phase by the waveshaper (Finisar 16000A) for optical arbitrary waveform generation \cite{shahverdi_quantum_2017, Jiang:07}. This can be controlled by a Matlab-based interface that observes the output by using an optical spectrum analyzer (OSA). The waveshaper provides pump (1551 nm) and signal (1559 nm) separately with designated temporal modes, verified via the frequency-resolved optical gating (FROG) method. The response time of the waveshaper is 500 ms. The pump and signal are further amplified by the two separated EDFAs to overcome the insertion loss $\sim4.5$ dB of the waveshaper. An optical delay line (ODL) in the pump arm is used to scan the relative time delay of the pump and signal pulses. Then polarization-controlled pump and signal pulses propagate in free space before being incident on two separate spatial-light modulators (Santec SLM-100, 1440×1050 pixels, pixel size 10×10 $\mu m^2$). The response time of both SLMs is 300 ms. We also added some free-space spectral filters to suppress the amplified spontaneous noise of the EDFA in the pump and signal arms. A beam splitter combines the signal and pump into the nonlinear crystal (5 mol.\% MgO doped PPLN crystal). We tested two separate nonlinear crystals with lengths 1 cm and 2.5 cm which generate sum frequency (SF) output photons with a normalized conversion efficiency of $\sim 1\%/W/cm^2$. Temporal walk-off between the pump and SF pulses for 1 cm and 2.5 cm long crystals are 1.2 ps and 3 ps, respectively. The parasitic pump and the generated second harmonic of the pump were filtered out, using two 850 nm long pass filters (FELH850) and three 775 nm bandpass filters (FBH770-10), with 100 dB and 150 dB suppression, respectively. After that, the SF output is coupled into single-mode fiber and detected by a high-speed single-pixel silicon avalanche photodiode (ID100). Then, a time-to-digital converter (TDC) can digitally count the number of photons and Matlab can post-process the data \cite{zhang_mode_2019}.

\begin{figure*}[!htb]
    \centering
    \includegraphics[width=1\linewidth]{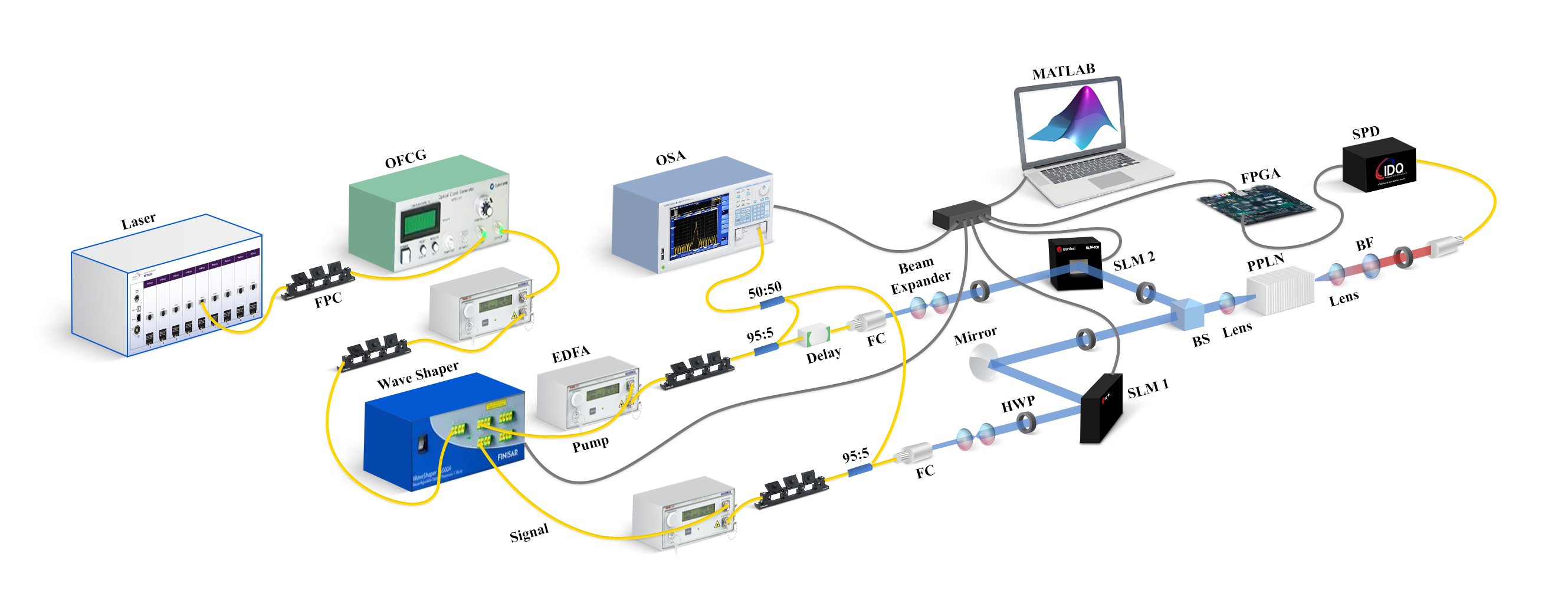}
    \caption{Experimental setup for programmable quantum parametric mode sorter in the ST domain. OFCG: Optical Frequency Comb generator, EDFA: Erbium Doped Fiber Amplifier, FPC: Fiber Polarization Controller, OSA: Optical Spectrum Analyzer, FC: Freespace to Fiber Coupler, HWP: Half Wave Plate, SLM: Spatial Light Modulator, BS: Beam-splitter, PPLN: Magnesium-doped Periodic Poled Lithium Niobate crystal, and SPD: Visible Single-Photon Detector, BF: Bandpass Filter.}
    \label{fig:Setup}
\end{figure*}

\section*{RESULTS}
\textbf{Effect of crystal length on selectivity:}
Due to limitations of the OFCG, the wavelengths of our pump and signal are only 8 nm apart. In this case, the inverse group velocity mismatch between the pump and signal can be neglected, making our system operate in the SSVM regime \cite{Reddy:13}. High mode selectivity is achieved when the temporal width of the pump is within the temporal walk-off between the pump and SF pulses in the crystal, i.e. the pump's spectral bandwidth exceeds the phase-matching bandwidth of the crystal. We plot the phase-matching curve of the crystal by sweeping the wavelength and measuring the generated SF power at a constant temperature. By fitting this experimental phase matching curve of the nonlinear crystal on top of an ideal $sinc^2$ curve, we evaluated the temporal walk-off between the pump and SF pulses in the 1 cm and 2.5 cm crystal to be 1.2 ps and 3 ps, respectively. In table \ref{table1}, we show the selectivity results for temporally optimized and un-optimized pump pulses for the 1 cm and 2.5 cm crystals (only for two temporal signal modes with spatial Gaussian). We chose a temporal pulse width of 2ps for both pump and signal pulses to distinctly show the relation between temporal walk-off and QPMS condition. We observe a larger selectivity for both $|T_0^p\rangle$ and $|T_1^p\rangle$ in the 2.5 cm crystal as compared to the 1 cm crystal because 2 ps is smaller than the temporal walk-off between the pump and SF pulses in the 2.5 cm crystal. We also show a significant improvement in the selectivity for both temporal modes by employing an optimized pump pulse.

\begin{table}
\begin{center}
\scalebox{0.5}{}
\begin{tabular}{|c||c|c||c|c|}
\hline
Crystal & $|T_0^p\rangle$ & Optimized & $|T_1^p\rangle$ & Optimized \\
length (cm) & (dB) & $|T_0^p\rangle$ (dB) & (dB) & $|T_1^p\rangle$ (dB)\\
\hline
1  & 6.26 dB  & 9.5 dB & 4.5 dB  & 4.5 dB \\
\hline
2.5  & 10.28 & 10.28 & 5.71 & 7.28 \\
\hline
\end{tabular}
\caption{\label{table1} Experimentally evaluated mode selectivity (using Eq. \ref{eq:ER}) between two signal modes $|T_{0}^{s}\rangle$ and $|T_{1}^{s}\rangle$ with 2ps pulse width, temporally optimized  and un-optimized pumps $|T_{0}^{p}\rangle$ and $|T_{1}^{p}\rangle$ for two different lengths of the crystals.}
\end{center}
\end{table}

In order to achieve higher selectivity in a 1 cm crystal, we have to use a temporal pulse width smaller than 1.2 ps. The spectral width of such a pulse is over 3 nm and spans near the peak of the phase matching for the second harmonic generation process, which leads to higher Raman noise. In order to significantly reduce the Raman noise, we use a temporal pulse width of over 2 ps with a spectral bandwidth of 1.6 nm through a 2.5 cm crystal. \newline

\textbf{Effect of pulse width on temporal selectivity:}
In table \ref{table2}, we show the selectivity results for 2 ps and 7 ps temporally optimized and un-optimized pump pulses with a constant spatial profile for both pump and signal as $|X_{0}^{p}, X_{0}^{s}\rangle$. In this case, 2 ps is less than the temporal walk-off between the pump and SF pulses in the crystal and yields higher selectivity than the 7 ps pulse. For instance, unoptimized mode $|T_0^p, T_0^s\rangle$, gives a $\sim 9.8$ dB difference in selectivity between 2 ps and 7 ps pulses. Even after significant improvement in selectivity by utilizing optimized pump pulses, we are only able to decrease this to $\sim6.4$ dB. In both cases, we observe that we can achieve high selectivity when the temporal shape of the pump and signal are identical $|T_{0}^{p}, T_{0}^{s}\rangle$, $|T_{1}^{p}, T_{1}^{s}\rangle$ and $|T_{2}^{p}, T_{2}^{s}\rangle$. $|T_{0}^{p}\rangle$ has a higher selectivity between $|T_{1}^{s}\rangle$ than $|T_{2}^{s}\rangle$, since $|T_{0}\rangle$ and $|T_{2}\rangle$ have a non-zero amplitude at zero delay. Our experimental results are on par with the theoretical simulations without the need for any fitting parameter. We also show how there is an abrupt drop in selectivity as the pulse width increases more than the temporal walk-off of the system in Appendix C. \newline
 
\begin{table*} \label{table:Mode 123}
\begin{center}
\scalebox{1}{
\begin{tabular}{|c||c|c|c|c||c|c|c|c|}
\hline
Pump Mode & Pulse Width & $|T_{0}^{s}\rangle$ & $|T_{1}^{s}\rangle$ & $|T_{2}^{s}\rangle$ & Pulse Width & $|T_{0}^{s}\rangle$ & $|T_{1}^{s}\rangle$ & $|T_{2}^{s}\rangle$ \\ & (ps) & (dB) & (dB) & (dB) & (ps) & (dB) & (dB) & (dB)\\
\hline
$|T_{0}^{p}\rangle$ & 2 & 10.28 & -11.24 & -17.97 & 7 & 0.42 & -8.34 & -2.73 \\
$|T_{1}^{p}\rangle$ & 2 & -10.96 & 5.71 & -7.98 & 7 & -3.75 & 3.06 & -14.53 \\
$|T_{2}^{p}\rangle$ & 2 & -6.5 & -7.66 & 3.11 & 7 & -9.06 & -4.63 & 2.37 \\
\hline
Optimized $|T_{0}^{p}\rangle$ & 2 & 10.28 & -16.39 & -17.97 & 7 & 3.87 & -13.93 & -4.72\\
Optimized $|T_{1}^{p}\rangle$ & 2 & -14.31 & 7.28 & -8.5 & 7 & -4.50 & 2.69 & -10.15\\
Optimized $|T_{2}^{p}\rangle$ & 2 & -21.43 & -6.76 & 6.55 & 7 & -8.74 & -11.29 & 6.38\\
\hline
\end{tabular}}
\caption{\label{table2}Experimentally evaluated temporal mode selectivity of the signal state $|T_{0}\rangle$, $|T_{1}\rangle$ and $|T_{2}\rangle$ with temporally optimized and un-optimized pump state $|T_{0}\rangle$, $|T_{1}\rangle$ and $|T_{2}\rangle$ for two different pulse widths (2 ps and 7 ps) in a 2.5 cm long nonlinear crystal. The selectivity values for each pump mode are calculated using Eq. \ref{eq:ER}.
}
\end{center}
\end{table*}
 
\begin{figure*}
    \centering
    \includegraphics[width=0.5\textwidth]{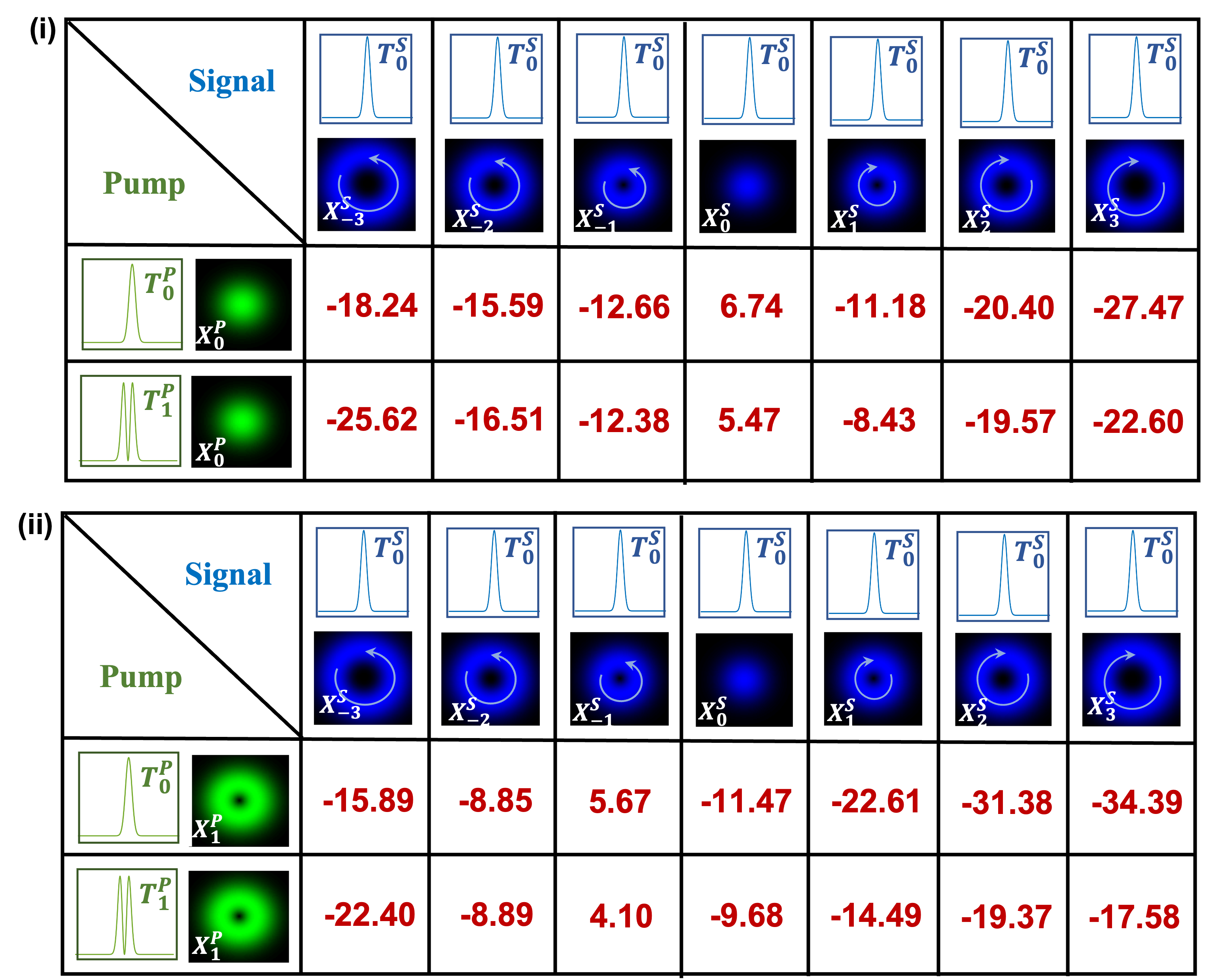}
    \caption{Selectivity results in dB measured experimentally for 5 different signal spatial modes and $|T_0^s\rangle$ temporal mode. Each row represents the pump in two different temporal modes ($|T_0^p\rangle$, $|T_1^p\rangle$) with spatial mode $|X_0^p\rangle$ in (i) and $|X_0^p\rangle$ in (ii).}
    \label{fig:SpatialS}
\end{figure*}

\textbf{Spatio-temporal mode selectivity:}
By combining spatial and temporal DoF, we take full advantage of the ST quantum states while being able to selectively upconvert a unique quantum state with high selectivity. Fig. \ref{fig:SpatialS} demonstrates how we can use different ST pump profiles to selectively up-convert a unique ST signal mode among the 7 modes considered. While $|X_0^p\rangle$ can efficiently up-convert a signal in $|X_0^s\rangle$ mode in Fig. \ref{fig:SpatialS}(i), we also show that $|T_0^p\rangle$ can upconvert a signal in $|T_0^s\rangle$ with higher selectivity than $|T_1^p\rangle$. Fig. \ref{fig:SpatialS} (ii) shows similar results for pump mode, $|X_1^p\rangle$ which can upconvert signal mode $|X_{-1}^s\rangle$ with high selectivity when their temporal modes are $|T_0^p, T_0^s\rangle$. We observe an $\sim$1.5 dB lower selectivity when the temporal profile of the pump and signal are not identical.

\begin{figure*}
    \centering
    \includegraphics[width=1\textwidth]{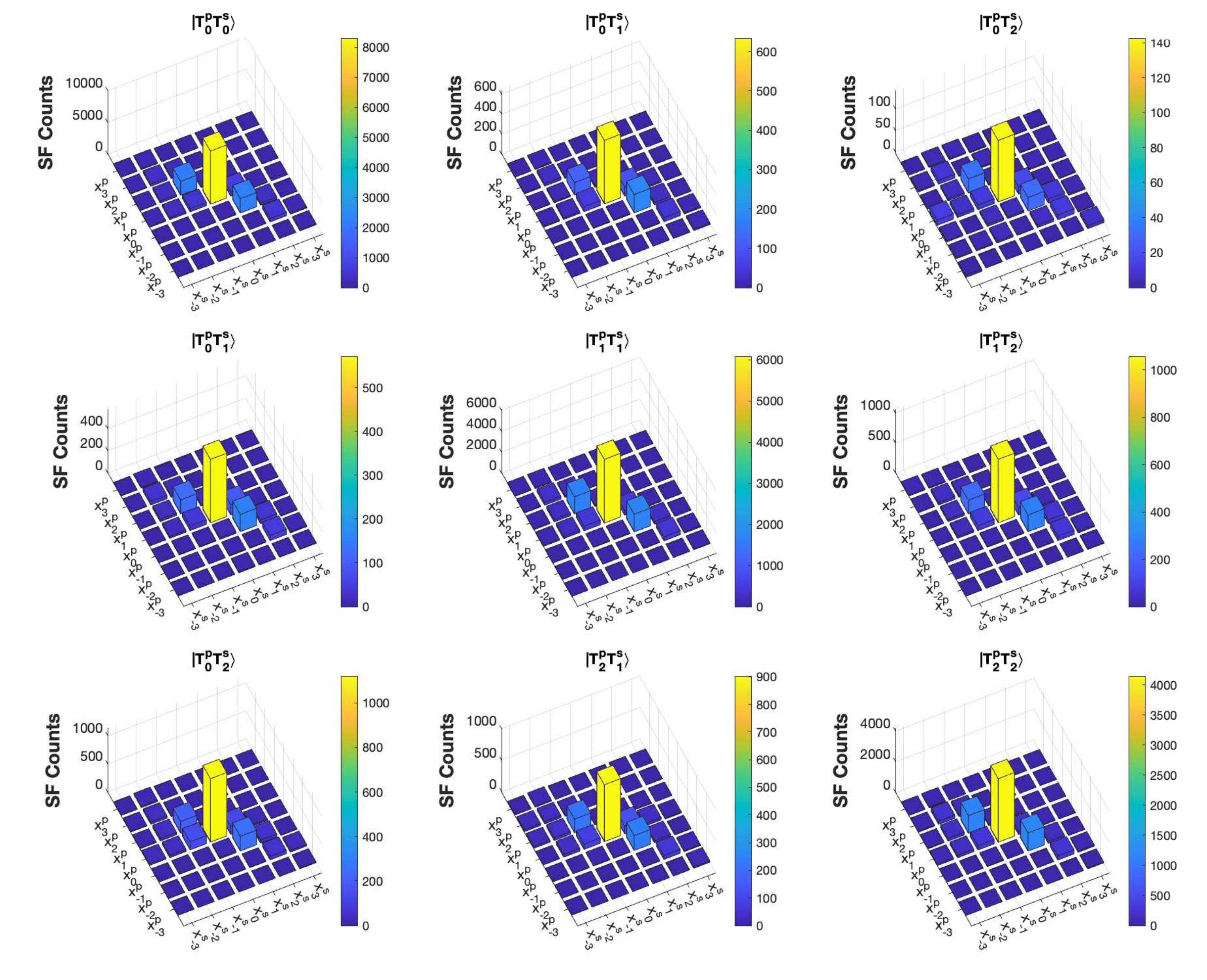}
    \caption{Measured SF photon counts with different combinations of ST signal and pump modes. Here, each of the 9 subfigures has a different combination of temporal HG pump and signal modes ($|T_{0}\rangle, |T_{1}\rangle$ and $|T_{2}\rangle$) with 2 ps pulse width. Each subfigure consists of SF counts data corresponding to 25 combinations of spatial LG pump and signal modes ($|X_{-2}\rangle, |X_{-1}\rangle, |X_{0}\rangle, |X_{1}\rangle$ and $|X_{2}\rangle$).}
    \label{fig:2psSFMatrix}
\end{figure*}

\begin{figure*}
    \centering
    \includegraphics[width=1\textwidth]{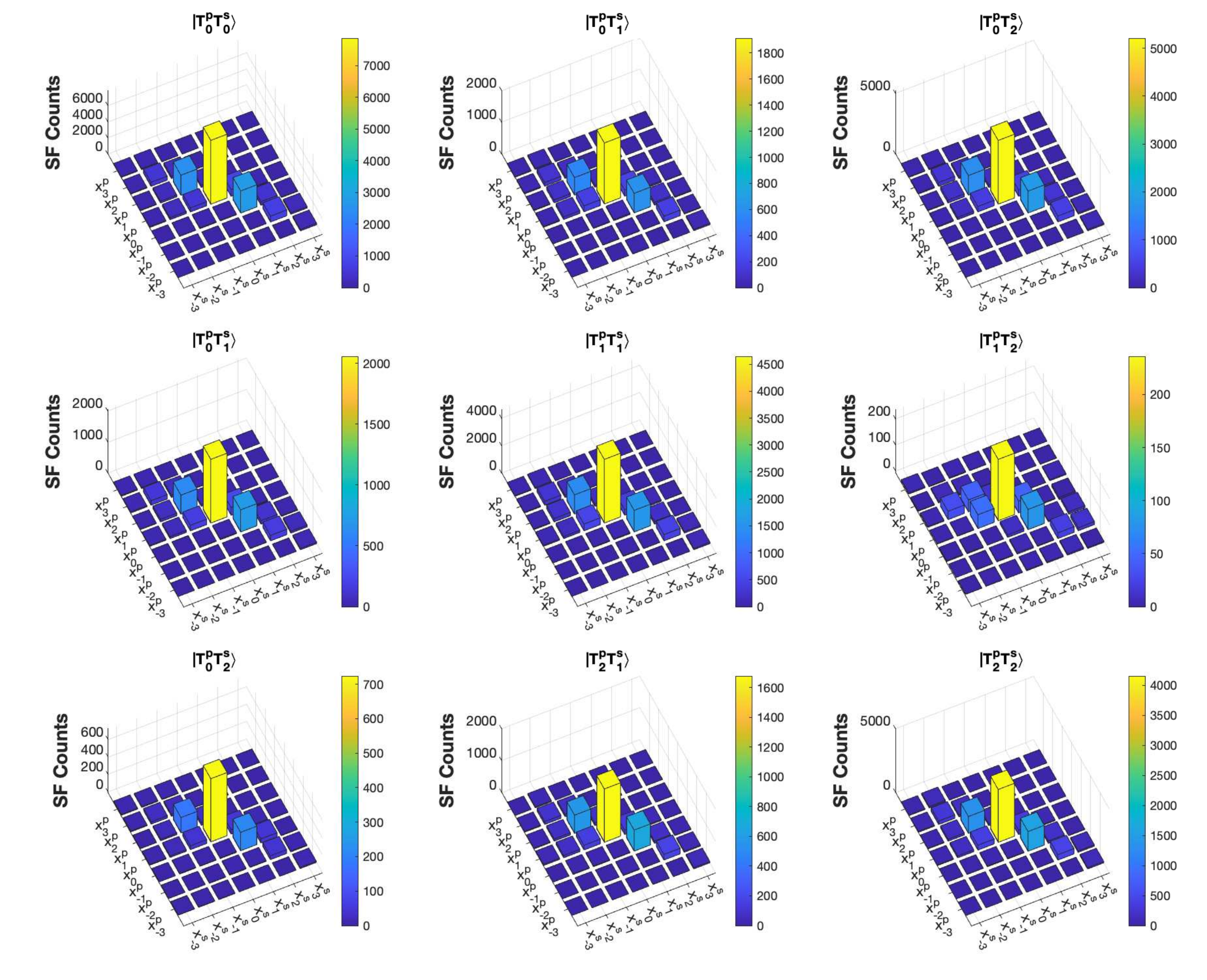}
    \caption{Measured SF photon counts with different combinations of ST signal and pump modes. Here, each of the 9 subfigures has a different combination of temporal HG pump and signal modes ($|T_{0}\rangle, |T_{1}\rangle$ and $|T_{2}\rangle$) with 7 ps pulse width. Each subfigure consists of SF counts data corresponding to 25 combinations of spatial LG pump and signal modes ($|X_{-2}\rangle, |X_{-1}\rangle, |X_{0}\rangle, |X_{1}\rangle,$ and $|X_{2}\rangle$).}
    \label{fig:7psSFMatrix}
\end{figure*}

In Fig. \ref{fig:2psSFMatrix} and \ref{fig:7psSFMatrix}, we show the tomography of spatial and temporal modes for 2 ps and 7 ps pulse width, respectively. Each of the 9 subplots represents the SF output for a different pair of temporal modes for pump and signal, respectively. Each row represents a different pump mode, whereas the columns represent different signal modes. For each combination of the pump and signal temporal mode, we observe the trend of SF photon counts for a combination of different spatial modes, ranging from $|X_{-2}\rangle$ through $|X_{2}\rangle$. Among the 9 subplots, we observe that the diagonal elements have the highest SF count as a consequence of having identical temporal modes. Also, within each subplot, we notice a similar trend where the diagonal elements which correspond to $|X_{i}\rangle|X_{-i}\rangle$ have a higher SF count as compared to the off-diagonal elements. \newline

\textbf{Selectivity of superposed modes:}
It has been shown that encoding information in higher dimensions like arbitrary super-positions of single photon temporal modes can significantly increase the security of the QKD protocol \cite{Manurkar:16, Habif:21, Tao:21}. Here, we consider two  additional modes in MUB by superposing two orthogonal temporal modes $|T_0\rangle$ and $|T_1\rangle$:

\begin{equation} \label{eq:T+}
    |T_+\rangle = \frac{1}{\sqrt{2}}(|T_{0}\rangle + |T_{1}\rangle)
\end{equation}
\begin{equation} \label{eq:T-}
    |T_-\rangle = \frac{1}{\sqrt{2}}(|T_{0}\rangle - |T_{1}\rangle)
\end{equation}
In table \ref{table3}, we show the selectivity results for these superposed modes by utilizing optimized and un-optimized pump pulses with a temporal width of 2 ps and 7 ps. It shows that we can pick individual superposed mode $|T^s_{+}\rangle$ ($|T^s_{-}\rangle$) against $|T^s_{-}\rangle$ ($|T^s_{+}\rangle$) and $|T^s_{2}\rangle$ modes with high selectivity. Similarly, $|T^s_{2}\rangle$ can be selected against $|T^s_{+}\rangle$ and $|T^s_{-}\rangle$. The selectivity can be further improved, in all three cases, by numerically optimizing the pump pulses. We use the PSO technique, as explained in the model section, to numerically prepare the optimized pump pulses. For 2ps temporal pulses, we observe an increase in selectivity from 22.71dB to 34.24dB for mode $|T_+^p\rangle$, 26.8dB to 28.05dB for mode $|T_-^p\rangle$, and 14.23dB to 19.17dB for mode $|T_2^p\rangle$. Similarly, for 7ps temporal pulses, the selectivity improves from 19.18dB to 27.07dB for mode $|T_+^p\rangle$, 22.08dB to 28.1dB for mode $|T_-^p\rangle$, and 10.69dB to 11.06dB for mode $|T_2^p\rangle$.

\begin{table*}
\begin{center}
\scalebox{1}{
\begin{tabular}{|c||c|c|c|c||c|c|c|c|}
\hline
Pump Mode & Pulse Width & $|T_{+}^{s}\rangle$ & $|T_{-}^{s}\rangle$ & $|T_{2}^{s}\rangle$ & Pulse Width & $|T_{+}^{s}\rangle$ & $|T_{-}^{s}\rangle$ & $|T_{2}^{s}\rangle$ \\ & (ps) & (dB) & (dB) & (dB) & (ps) & (dB) & (dB) & (dB)\\
\hline
 $|T_{+}^{p}\rangle$ & 2 & 9.17 & -13.54 & -11.54 & 7 & 7.81 & -11.37 & -10.98 \\
 $|T_{-}^{p}\rangle$ & 2 & -16.20 & 10.61 & -12.23 & 7 & -13.43 & 8.65 & -10.81 \\
$|T_{2}^{p}\rangle$ & 2 & -8.17 & -9.19 & 5.04 & 7 & -7.47 & -7.23 & 3.46 \\
\hline
Optimized $|T_{+}^{p}\rangle$ & 2 & 13.41 & -20.83 & -14.35 & 7 & 10.16 & -16.91 & -11.37 \\
Optimized $|T_{-}^{p}\rangle$ & 2 & -16.38 & 11.67 & -13.67 & 7 & -17.31 & 10.79 & -12.06 \\
Optimized $|T_{2}^{p}\rangle$ & 2 & -10.74 & -11.45 & 7.72 & 7 & -8.31 & -5.54 & 2.75 \\
\hline
\end{tabular}}
\caption{\label{table3} Experimentally recorded temporal mode selectivity results using superposed signal modes $|T_{+}^{s}\rangle$, $|T_{-}^{s}\rangle$ and mode 3 ($|T_{2}^{s}\rangle$) with temporally optimized and un-optimized pump mode $|T_{+}^{p}\rangle$, $|T_{-}^{p}\rangle$ and mode 3 ($|T_{2}^{p}\rangle$) for two different pulse widths in a 2.5 cm long crystal. The selectivity values are similarly calculated using Eq. \ref{eq:ER}.}
\end{center}
\end{table*}

\section*{DISCUSSION}
In principle, ST quantum states can form an infinite Hilbert space, which increases the information capacity of a single photon. Convenience in encoding and decoding quantum information in these ST Hilbert spaces can make it beneficial for achieving enhanced key rates for HD quantum key distribution. In this work, we prepare quantum states in a unique set of spatial and temporal modes by using a combination of an SLM, an OFCG, and a waveshaper. We demonstrate a programmable quantum parametric mode sorter that can pick the desired ST mode with high selectivity using the SF process in a $\chi^2$ nonlinear crystal. We are able to achieve between $12.99$ dB to $19.4$ dB selectivity among two overlapping spatial modes and between $10.77$dB to $21.52$ dB selectivity among two overlapping temporal modes. We can further improve this up to $\sim 26.67$dB by optimizing temporal pump modes.

In our system, there are some technical limitations. For instance, in generating higher-order spatial and temporal quantum states. To create ultrafast pulses in the temporal domain, we use OFCG with a line spacing of 25 GHz (spectral spacing 0.2 nm) limits the effective number of comb lines used for pulse shaping. For a 2 ps pulse, the spectral FWHM is 1.6 nm, which means that we effectively have 8 comb lines in the 3 dB bandwidth range, this is inadequate to realize modes higher than third order. Similarly, the higher-order modes in the spatial domain with opposite orbital angular momentum for signal and pump give low photon counts after coupling into the single mode fiber ($\sim$4 $\mu$m core diameter). This can be seen in the diagonal elements of Fig. \ref{fig:2psSFMatrix} and \ref{fig:7psSFMatrix}. Replacing single-mode fiber with few-mode fiber could help us couple higher-order OAM quantum states \cite{PhysRevApplied.11.064058, Roadmap} but the spatial selectivity would be compromised.

In the current setup, we generate ultra-fast picosecond level temporal pulses by individually manipulating the phase of 37 optical frequency comb lines using a Finisar waveshaper. We found that when the phase patterns are refreshed in the waveshaper, it introduces a temporal jitter of $\sim$ 2 ps, which can disturb the temporal overlap between pump and signal input pulses within the crystal. We overcame this issue by scanning the SF counts for the different delays between the pump and the signal using an optical delay line. Since this is in excellent agreement with the simulation results, we are able to overlap them and measure the counts at the complete intersections between the input pulses. Our delay and SF count measurement for each pair of pump and signal modes roughly take about 1 second in our Matlab-based interface. This, though fast enough for the initial assembling of the system, could be a substantial hindrance when it comes to adaptively optimizing the temporal shape of the pump by using the SF photon counts as feedback. However, it may be possible to achieve a higher selectivity by adaptively optimizing the temporal pump pulse at the cost of increased operation time.

An on-demand single photon source is vital for most quantum technological applications because of its negligibly low jitter, high spectral purity, and high collection efficiency. In quantum optics experiments, spontaneous parametric down-conversion is commonly used as a heralded single photons source, since we can tailor the joint spectral amplitude function to control the temporal mode structure of the generated photon pair \cite{Ansari:18, Ansari:20}. However, multi-photon pair generation reduces the purity of the single photons and can lead to errors in the operation of quantum circuits. One way to reduce the generation of these multi-photons is by weakly pumping the nonlinear crystal, on the other hand, reduces the mean photon number and leads to a lower probability of single photon output. We can overcome this inherent inefficiency of the heralded single photons produced in an SPDC crystal by selectively up-converting the output heralded and heralding single photons by using two identical quantum parametric mode sorters, as described in this paper, to achieve a single photon source with a highly diminished mode-mismatch. We can also use photon number resolving detectors to monitor the number of photon pairs per pulse and select only those cases for which only one pair is generated, and further ensure the purity of the shaped single photons \cite{Kiyohara:16, Bonneau_2015}.
To beneficially utilize the enormous information capacity of a single photon in an HD ST Hilbert space, individual ST modes can be extracted by cascading the current QPMS scheme in a serial or loop structure \cite{PhysRevLett.106.120403,huang_mode-resolved_2013, PhysRevA.98.023836, Kaneda:20}. We can further optimize the temporal and spatial tomography of the pump in the present QPMS scheme to mitigate both, the effect of atmospheric turbulence on the OAM states \cite{Br_nner_2013, Tao:21, zhang_mode_2019} and the effect of broadband noise overlapping in the spectral and temporal domain \cite{shahverdi_quantum_2017}.

\section*{CONCLUSION}
In summary, we have developed a fully programmable quantum parametric mode sorter on a combined ST mode basis. By utilizing the optimized pump modes, coupling into the single-mode fiber, and operating at the edge of the phase-matching in a nonlinear crystal, we showed the effective mode-sorting performance in the HD ST Hilbert space. Only optimizing pumps in the temporal domain can achieve more than 12 dB extinction for MUB sets of ST modes.
This technique could find applications in HD quantum communications and computations, quantum cryptography, pattern recognition, and quantum LiDAR \cite{PhysRevX.10.031031, Rehain2020, Shahverdi:18, Zhu:21}, and so on. In the future, we can extend our HD quantum mode-sorting approach for more practical scenarios, such as passing through a noisy environment due to spatial turbulence \cite{zhang_mode_2019} and/or time-frequency noises \cite{shahverdi_quantum_2017} which could overlap with the desired signal modes.

\newpage

\section*{APPENDIX A} 
\begin{figure*}[!htb]
    \centering
    \includegraphics[width=1\textwidth]{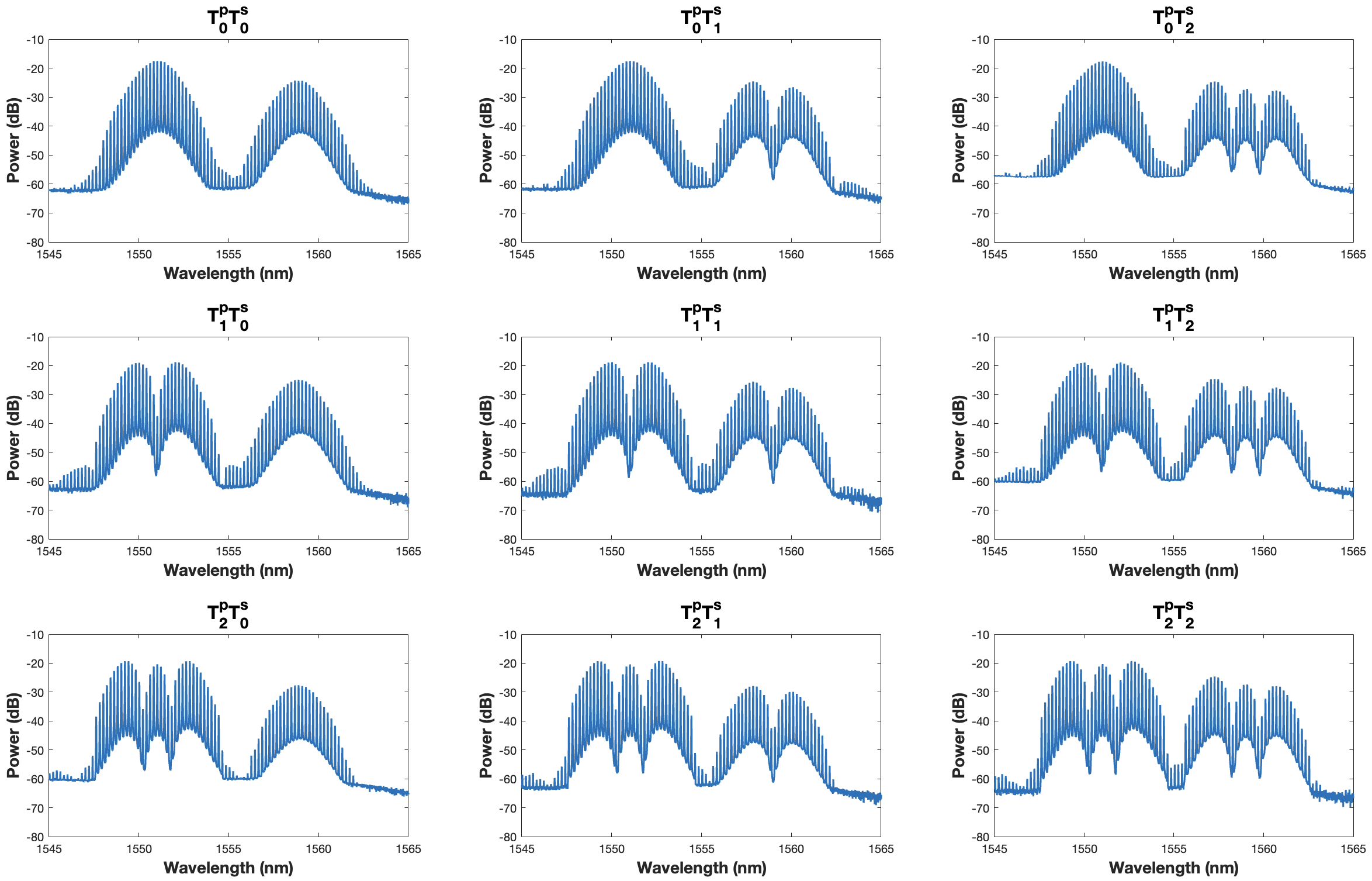}
    \caption{2ps temporal pump and signal profiles in the spectral domain as read by an optical spectrum analyzer after 13dB attenuation. The central wavelength of the pump and signal are 1551 nm and 1559 nm, respectively.}
    \label{fig:Spectral2ps}
\end{figure*}

\textcolor{blue}{\textit{\textbf{Spectral profiles of 2 ps temporal pulses:}}}
We generate ultra-fast temporal pulses by individually manipulating the phase and amplitude of optical frequency comb lines generated by an OFCG. We combine 5\% of pump and signal pulses using a beam combiner to observe them in the spectral domain using an OSA. In Fig. \ref{fig:Spectral2ps}, we show the spectral profiles of different modes of pump and signal pulses with center wavelengths 1551 nm and 1559 nm. We observe a slight power at the phase-matching wavelength for second harmonic generation ($\sim$1555 nm), which, in a single photon regime, gives rise to considerable noise photon counts. To avoid this, we further make use of free space short pass and long pass filters on the signal and pump arm, respectively.

\section*{APPENDIX B} 
\begin{figure*}[!htb]
    \centering
    \includegraphics[width=1\linewidth]{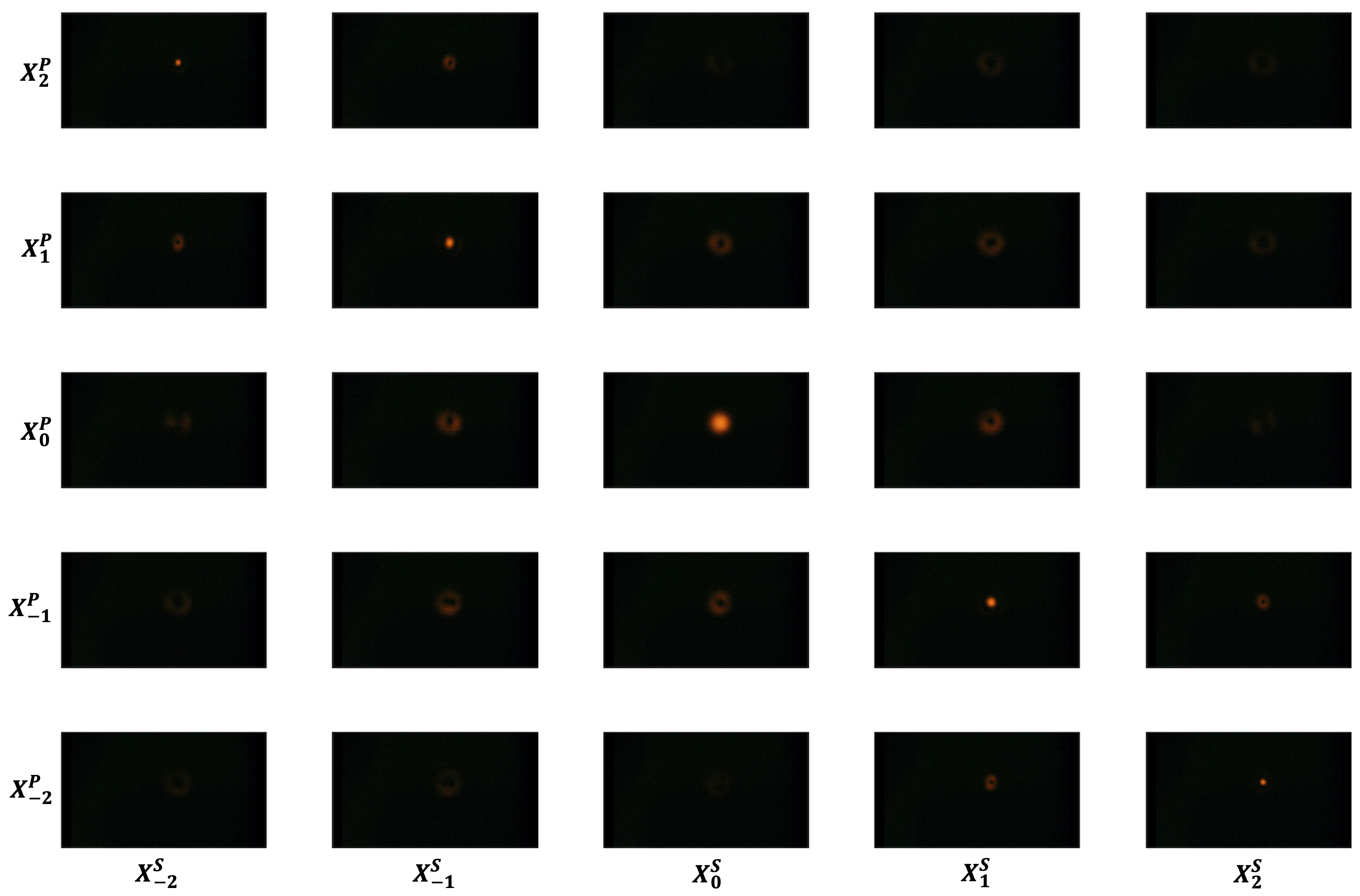}
    \caption{Experimentally measured spatial profiles of SF output, detected on a CCD camera (for classical measurement), using different sets of signal and pump modes. }
    \label{fig:CameraImage}
\end{figure*}

\textcolor{blue}{\textit{\textbf{Camera image of spatial modes:}}}
To verify the profile of the spatial modes, we experimentally capture the classical SF output on a CCD camera as shown in \ref{fig:CameraImage}. The spatial intensity profile of the SF output for the combination of $|X^p_{-2}\rangle$, $|X^p_{-1}\rangle$, $|X^p_0\rangle$, $|X^p_1\rangle$, and $|X^p_2\rangle$ pump modes with $|X^s_{-2}\rangle$, $|X^s_{-1}\rangle$, $|X^s_0\rangle$, $|X^s_1\rangle$, and $|X^s_2\rangle$ signal modes are in excellent agreement with the simulated results shown in Fig. \ref{fig:Model}(v). 

\section*{APPENDIX C} \textcolor{blue}{\textit{\textbf{QPMS and temporal walk-off:}}}
Table \ref{table4}, illustrates how mode selectivity is influenced by the temporal walk-off between the pump and SF pulses in the crystal. Signal modes $|T_0\rangle$ and $|T_1\rangle$ are selectively up-converted using a temporally optimized and unoptimized pump pulse in a 1 cm crystal. We show a sudden drop in the selectivity beyond a temporal pulse width of 1 ps since the temporal walk-off between the pump and SF pulses in the crystal is $\sim$1.2 ps. Though there is a drop in selectivity between 2 ps and 3 ps pulses, they are comparable to each other. \newline \newline

\begin{table}
\begin{center}
\scalebox{0.97}{
\begin{tabular}{|c||c|c||c|c| }
\hline
Pulse width & $|T_0^p\rangle$ & Optimized $|T_0^p\rangle$ & $|T_1^p\rangle$ & Optimized $|T_1^p\rangle$ \\ (ps) & (dB) & (dB) & (dB) & (dB)\\
\hline
1  & 8.03 & 11.1 & 7.6  & 8.1  \\
\hline
2 & 6.26  & 9.5 dB & 4.5 & 4.5 \\
\hline
3 & 5.0  & 8.2  & 2.74  & 4.4 \\
\hline
\end{tabular}}
\caption{\label{table4} Three different temporal pulse widths are used to experimentally evaluate the mode selection of signal mode 1 ($HG_{00}$) and 2 ($HG_{01}$) with temporally optimized and un-optimized pumps using 1 cm crystal.}
\end{center}

\end{table}

\noindent\textbf{Funding:} This research was supported by the NSF and Earth Science Technology Office, NASA.

\noindent\textbf{Disclosures.} The authors declare no conflicts of interest.

\noindent\textbf{Acknowledgment.} The authors thank Jeevanandha Ramanathan for helping with the FPGA-based photon counting system and Bhavya Mohan for helping with the 3D diagram of the experimental setup.


\bibliography{Ref}

\end{document}